\documentclass[a4paper,11pt]{article}
\usepackage{jheppub}
\usepackage[utf8]{inputenc}
\usepackage{mathtools}
\usepackage{caption}
\usepackage{subcaption}

\def \L {\mathcal{L}}
\newcommand{\dd}{{\rm d}}
\newcommand{\ii}{{\rm i}}

\title{Gauged Q-balls in flat potentials}
\author[1]{Julian Heeck,%
\note{ORCID: \href{https://orcid.org/0000-0003-2653-5962}{0000-0003-2653-5962}}
}   
\author[2]{Yu Zhi 
\note{ORCID: \href{https://orcid.org/0009-0007-3440-4678}{0009-0007-3440-4678}}
}
\affiliation{Department of Physics, University of Virginia, 
Charlottesville, Virginia 22904-4714, USA}

\emailAdd{heeck@virginia.edu}
\emailAdd{yz9hy@virginia.edu}

\abstract{
Q-balls are large bound-state systems of scalar particles, described classically through localized solutions of the equations of motion. Promoting the required stabilizing $U(1)$ symmetry to a gauge symmetry leads to gauged Q-balls, which cannot grow beyond some maximal size and charge on account of the repulsive gauge interactions. These gauged Q-balls have been studied extensively for scalar potentials that satisfy Coleman's thin-wall criterion; here, we explore gauged Q-balls in flat potentials, which often occur in supersymmetric models. Even though global Q-balls in flat potentials are qualitatively different from Coleman's Q-balls, we find that the gauged versions are remarkably similar. We provide analytic approximations for these solitons and compare to numerical solutions. In addition, we study Proca Q-balls, i.e.~make the gauge bosons massive, which interpolates between the global and gauged cases.
}

\begin{document}
\maketitle
\flushbottom

\section{Introduction}

Non-topological solitons are localized scalar field configurations held together by attractive interactions and stabilized by some conserved Noether charge~\cite{Zhou:2024mea}. Coleman's Q-balls~\cite{Coleman:1985ki} are simple examples, requiring only a single complex scalar field $\phi(\vec{x},t)$ with a particular $U(1)$-symmetric potential $U(|\phi|)$. Q-ball solitons then arise as spherically-symmetric solutions to the classical Euler--Lagrange equations, which can be the lowest-energy configuration of $Q$ scalars, thus rendering them stable. Coleman's Q-balls have mass or energy that is proportional to the number of scalars inside, $E\propto Q$, at least in the limit of large $Q$, which allows for simple analytic approximations in the so-called thin-wall regime~\cite{Coleman:1985ki,Heeck:2020bau,Heeck:2022iky,Aiello:2026jay}. However, these are not the only type of non-topological solitons. For \textit{flat potentials}, Coleman's analysis is not applicable, but one can still find stable localized scalar field configurations, which actually predate Coleman's solutions by decades~\cite{Rosen:1968mfz}. These objects -- still called Q-balls here -- are more diffuse and have a different energy scaling for large $Q$, $E\propto Q^{3/4}$~\cite{Dvali:1997qv}, and are thus inevitably stable for large-enough $Q$.

Neither of these Q-ball types arise in renormalizable single-field  potentials, but both can be realized in certain limits of renormalizable \textit{multi-field} theories, see for example Ref.~\cite{Heeck:2022iky} for UV-complete realizations of Coleman-type Q-balls, and the Friedberg--Lee--Sirlin model~\cite{Friedberg:1976me} as a realization of flat-potential Q-balls~\cite{Heeck:2023idx}. As observed long ago, supersymmetric extensions of the Standard Model of particle physics generically contain many flat directions~\cite{Gherghetta:1995dv,Enqvist:2003gh} that could lead to these kind of solitons~\cite{Kusenko:1997zq,Kusenko:1997si,Zhou:2024mea}, with baryon and lepton number playing the role of the stabilizing $U(1)$ symmetries. This observation has led to significant interest in these types of solitons, given that they arise generically in well-motivated theories and could then form dark matter~\cite{Dvali:1997qv}.

The soliton description along flat directions in vast supersymmetric potentials is usually approximated as a single-field problem with an effective potential~\cite{Dvali:1997qv}. The particle-physics picture would be a large bound state of squarks and sleptons, in just the right ratios to keep the Q-ball electrically neutral, at least if the goal is to describe dark matter. Inside the Q-ball, these sparticles however not only interact through the scalar potential, but also via gauge interactions, seeing as they carry the same charges as quarks and leptons. Additional gauge interactions arise if the global $U(1)_{B-L}$ is promoted to a gauge symmetry, often done to explain $R$ parity, i.e.~eliminate proton decay and render the lightest superparticle stable~\cite{Martin:1992mq}. 
Gauge interactions complicate the soliton description and might even qualitatively change it: \textit{gauged} Q-balls~\cite{Rosen:1968zwl,Lee:1988ag}, i.e.~Q-balls based on a Lagrangian with \textit{local} $U(1)$ symmetry, cannot grow indefinitely before the repulsive gauge interactions render them unstable. The resulting \textit{maximal} charge $Q$ could be in contradiction with the \textit{minimal} charge these Q-balls are required to have to forbid decay into Standard-Model fermions~\cite{Kusenko:1997si,Campanelli:2007um}, which would eliminate these Q-balls as dark matter.

While a full analysis of supersymmetric Q-balls in flat potentials including gauge interactions is beyond the scope of this article, we take a first step in this direction by studying gauged Q-ball in flat potentials. Gauged Q-balls in Coleman-like potentials have been discussed for almost forty years~\cite{Lee:1988ag} and can be understood well analytically~\cite{Heeck:2021zvk}. An analogous discussion for flat potentials is missing in the literature to the best of our knowledge, despite the intense interest in flat-potential Q-balls as dark matter.
Some related work can be found in Refs.~\cite{Gulamov:2013cra,Hong:2015wga,Nugaev:2017lls}; notably,   Refs.~\cite{Lee:1991bn,Loiko:2019gwk,Loiko:2022noq} study gauge interactions in the Friedberg--Lee--Sirlin model, but do not provide good descriptions for large solitons. Finally, Ref.~\cite{Brihaye:2014gua} provides a numerical analysis of gauged Q-balls in flat potentials -- even including gravitational effects -- but without analytic approximations.

The rest of this article is organized as follows: we discuss the familiar case of global Q-balls in flat potentials in Sec.~\ref{sec:global} to introduce our notation. The gauged case is discussed in Sec.~\ref{sec:gauged}, both numerically and via analytic approximations in the large Q-ball limit. In Sec.~\ref{sec:proca}, we introduce a mass for the gauge boson, leading to Proca Q-balls that contain global and gauged Q-balls as limiting cases but exhibit novel features. Finally, we conclude in Sec.~\ref{sec:conclusions}.
Data tables of our numerical solutions are available as ancillary files of the arXiv version of this article.

\section{Global Q-Balls}
\label{sec:global}

\begin{figure}[t]
    \centering
    \includegraphics[width=0.6\textwidth]{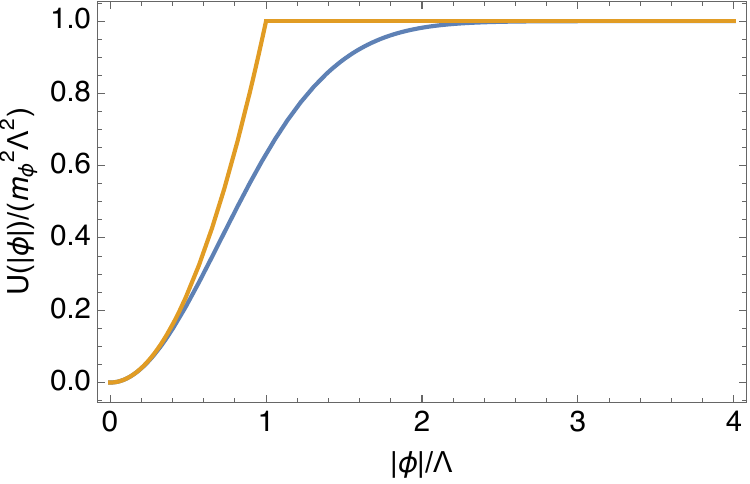}
    \caption{The flat potential of Eq.~\eqref{eq:Uexact} (blue) and its approximation of Eq.~\eqref{eq:Uapprox} (orange).}
     \label{fig:UGlobal}
\end{figure}

To set the stage, we revisit the well-known case of \textit{global} Q-balls in flat potentials~\cite{Rosen:1968mfz,Dvali:1997qv,Campanelli:2007um,Copeland:2009as,Espinosa:2023osv}. We assume a complex scalar field $\phi$ with a Lagrangian
\begin{align}
    \L = |\partial_\mu \phi|^2 - U(|\phi|) 
\end{align}
that is invariant under a \emph{global} $U(1)$ symmetry $\phi \left(\vec{x},t\right) \to e^{\ii \alpha} \phi \left(\vec{x},t\right) $. We take the potential $U (|\phi|)$ to be~\cite{Rosen:1968mfz,Copeland:2009as}
\begin{align}
     U (|\phi|) = m^2_\phi \Lambda^2 \left(1 - e^{-\frac{|\phi|^2}{\Lambda^2}}\right),
     \label{eq:Uexact}
\end{align}
which is a smoothed-out version of the simple piecewise-quadratic potential
\begin{align}
    U(|\phi|) =
    \begin{cases}
    m_\phi^2 |\phi|^2 \,, & \text{ for }  |\phi| \leq \Lambda \,,\\[6pt]
    m_\phi^2 \Lambda^2\,, & \text{ for } |\phi| > \Lambda \,,
    \end{cases}
    \label{eq:Uapprox}
\end{align}
with $\phi$ mass $m_\phi$ and some scale $\Lambda$ that determines when the potential becomes flat, see Fig.~\ref{fig:UGlobal} for a visualization. While we take Eq.~\eqref{eq:Uexact} as the potential for concreteness in the following, we expect our results to approximately apply to other flat potentials, as long as they reduce to $|\phi|^2$ for small $\phi$ and to something that grows \textit{slower} than $|\phi|^2$ for large $\phi$.

We search for solutions to the Euler--Lagrange equations of $\L$ in the form of a localized spherically-symmetric configuration with constant-phase time dependence,
\begin{align}
    \phi\left(\vec{x},t\right) = \Lambda e^{\ii\omega t}  f(r) \,, 
\end{align}
where $r\equiv |\vec{x}|$ and $f(r)$ is the real-valued dimensionless profile function of interest, essentially setting the $\phi$ density inside the Q-ball. $\omega$ is a free parameter that determines the Q-ball radius, charge, and energy, and can also be viewed as the chemical potential because $ \dd E/\dd Q=\omega$~\cite{Zhou:2024mea}. For convenience, we define the dimensionless quantities
\begin{align}
    \kappa \equiv \frac{\omega}{m_\phi}\,, \quad \text{ and } \quad \rho \equiv r\  m_{\phi} \,,
\end{align}
leading to the Euler--Lagrange differential equation for $f(\rho)$,
\begin{align}
  \boxed{  f'' + \frac{2}{\rho}f' = fe^{-f^2} - \kappa^2 f \,, }
    \label{eq:global_diffeq}
\end{align}
which is to be solved subject to the boundary conditions
\begin{align}
    \lim_{\rho \rightarrow 0}f' = \lim_{\rho \rightarrow \infty} f = 0
\end{align}
to ensure a localized solution. The parameter $\kappa$ is restricted to be between 0 and 1. We further restrict ourselves to monotonically decreasing solutions, which correspond to the Q-ball ground states~\cite{Volkov:2002aj,Mai:2012cx,Almumin:2021gax}.
Following Coleman, the differential equation can be interpreted as a one-dimensional classical-mechanics problem, in which the position $f$ of a particle changes with time $\rho$ due to the friction term $2f'/\rho$ and the potential
\begin{align}
    V (f)= \frac{1}{2}\kappa^2 f^2 - \frac{1}{2}\left(1 - e^{-f^2}\right) .
\end{align}
The particle starts at rest and rolls up to the local maximum at $f=0$. Unlike the potentials discussed by Coleman~\cite{Coleman:1985ki}, $V$ does not feature another maximum and his step-function thin-wall approximation is not applicable.

Numerical solutions to Eq.~\eqref{eq:global_diffeq} can still be found using Coleman's shooting method.\footnote{Many computer codes such as \texttt{AnyBubble}~\cite{Masoumi:2016wot} -- designed to solve structurally identical differential equations for vacuum decay~\cite{Coleman:1977py} -- cannot be used for flat potentials because they correspond to tunneling into an unbounded region.}
We employed a different method based on finite differences, for which we first switch to a new variable
 $y \equiv \rho/(1+\frac{\rho}{a})$~\cite{Heeck:2020bau}, where $a$ is a  large finite number and rewrite the differential equation in terms of $y$:
\begin{align}
    \left( 1 - \frac{y}{a}\right)^4 \left( f'' + \frac{2}{y}f'\right) + f\left(\kappa^2 - e^{-f^2}\right) = 0 \,.
    \label{eq:global_diffeq_y}
\end{align}
The boundary conditions are then in a finite domain, far easier to implement numerically:
\begin{align}
    \lim_{y \rightarrow 0}f' = \lim_{y \rightarrow a} f = 0\,.
\end{align}
We compared our \texttt{Mathematica} \texttt{NDSolve} solution of Eq.~\eqref{eq:global_diffeq_y} with shooting solutions and found good agreement, up to numerical noise. Our finite-element solver requires a test function for $f(\rho)$ as input, which we construct as follows.

\begin{figure}[t]
    \centering
    \begin{subfigure}{0.48\textwidth}
        \includegraphics[width=\linewidth]{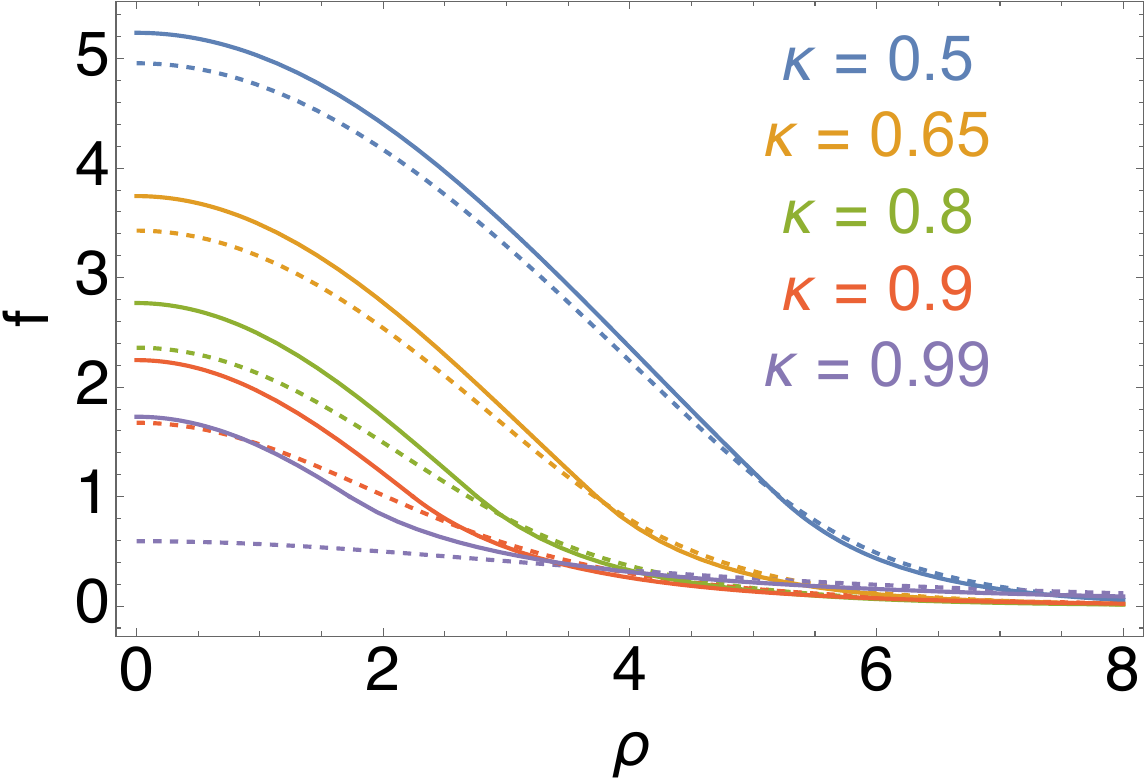}
    \end{subfigure}
    \hfill
     \begin{subfigure}{0.5\textwidth}
        \includegraphics[width=\linewidth]{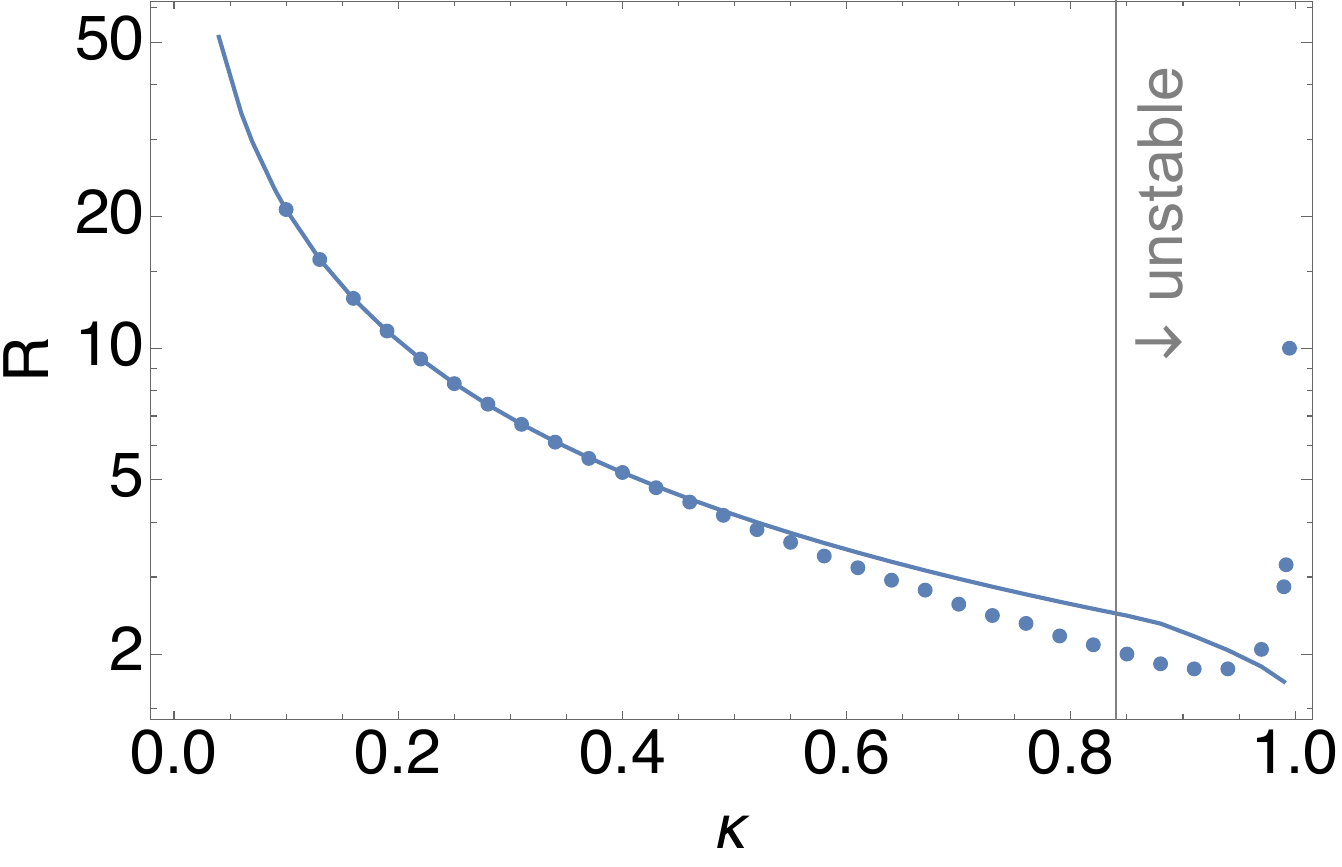}
    \end{subfigure}
    \caption{Left: Profiles $f(\rho)$ for different $\kappa$ values. Dashed lines are numerical solutions, solid lines are analytical predictions from Eq.~(\ref{fPG}).
    Right: The global Q-ball radius $R$ vs $\kappa$. The solid line is the analytical prediction of the global Q-ball radius solved from Eq.~\eqref{fPG}; points are numerical solutions. The region $\kappa > 0.84$ leads to unstable Q-balls.
    }
     \label{fig:fRGlobal}
\end{figure}

Approximating the potential as in  Eq.~\eqref{eq:Uapprox} allows for an analytical solution of the piecewise-linear differential equation,
\begin{equation}
    f(\rho) =
\begin{cases}
f(0)\frac{\sin(\kappa \rho)}{\kappa \rho}\,, & \rho \leq \rho^* \,,\\[6pt]
f(0)\frac{\sin(\kappa \rho^*)}{\kappa \rho} e^{\sqrt{1-\kappa^2}(\rho^*-\rho)}\,, & \rho > \rho^* \,,
\end{cases}
\label{fPG}
\end{equation}
with connection point $\rho^*$  given by 
\begin{align}
    \rho^* \equiv \frac{\pi - \arcsin \kappa}{\kappa} = \frac{\pi}{\kappa} -1 - \mathcal{O}(\kappa^2)\,.
    \label{eq:rho_star}
\end{align}
The prefactor $f(0)$ cannot be fixed by the linear differential equation or the boundary conditions, but can be obtained as in Ref.~\cite{Espinosa:2023osv}. First off, the global Q-ball charge $Q$ and energy or mass $E$ are defined through the integrals
\begin{align}
    & Q = 8\pi \kappa \frac{\Lambda^2}{m_\phi^2} \int \text{d}\rho \rho^2f^2\,, \\
    & E = \kappa m_\phi Q+\frac{8\pi}{3} \frac{\Lambda^2}{m_\phi}\int \text{d}\rho \rho^2 {f'}^2\,,
\end{align}
using a virial-theorem relation to rewrite $E$~\cite{Derrick:1964ww,Friedberg:1976me}. These macroscopic quantities satisfy the non-trivial differential relationship~\cite{Zhou:2024mea}
\begin{align}
    \frac{\text{d}E}{\text{d}\omega} = \omega\frac{\text{d}Q}{\text{d}\omega} \quad \Leftrightarrow \quad  \int\dd\rho \rho^2 f^2 = -\frac{1}{3\kappa}\frac{\dd}{\dd\kappa} \int\dd\rho \rho^2 {f'}^2
    \label{dEdQRG}
\end{align}
if $f$ solves the equation of motion. With our ansatz from Eq.~\eqref{fPG}, the  two essential integrals take the form~\cite{Espinosa:2023osv}
\begin{align}
    &\int \text{d}\rho \rho^2f^2 = \frac{f(0)^2}{4\kappa^3} \left( 2\cos^{-1}(\kappa) + \pi + \frac{2\kappa}{\sqrt{1-\kappa^2}}\right) ,
    \label{QintG}\\
    &\int \text{d}\rho \rho^2f^{\prime 2} = \frac{f(0)^2}{4\kappa} \left( 2\cos^{-1}(\kappa) + \pi\right),
    \label{EintG}
\end{align}
which satisfy Eq.~\eqref{dEdQRG} if the prefactor is set to $f(0) = \rho^*$ from Eq.~\eqref{eq:rho_star}, using additional arguments to fix an overall constant~\cite{Espinosa:2023osv}. With this, we have a complete analytical approximation for the Q-ball profile, which works especially well for small $\kappa$, see Fig.~\ref{fig:fRGlobal} (left), and is also a useful test function for our numerical solver.
It is worth emphasizing that unlike Coleman's Q-ball profiles, the flat-potential solutions are never constant inside the Q-ball, leading to a more diffuse object. Nevertheless, these Q-balls also grow larger and larger for decreasing $\kappa$, requiring larger and larger initial field values, $f(0)\simeq \pi/\kappa$.

Here and in the following, we define the Q-ball radius $R$ via $f''(R) = 0$. For small $\kappa$, we approximately have
\begin{align}
    R \simeq \frac{2}{3} \rho^* \simeq \frac{2\pi}{3\kappa} \,,
\end{align}
so the small-$\kappa$ limit indeed corresponds to the large Q-ball limit.
The full relationship between radius and $\kappa$ is shown in Fig.~\ref{fig:fRGlobal} (right), together with the prediction from our ansatz~\eqref{fPG}. We can see clearly that our ansatz is excellent for $\kappa \ll 1/2$, but fails to describe the behavior near $\kappa \sim 1$.

\begin{figure}[t]
  \centering
  \begin{subfigure}{0.48\textwidth}
    \includegraphics[width=\linewidth]{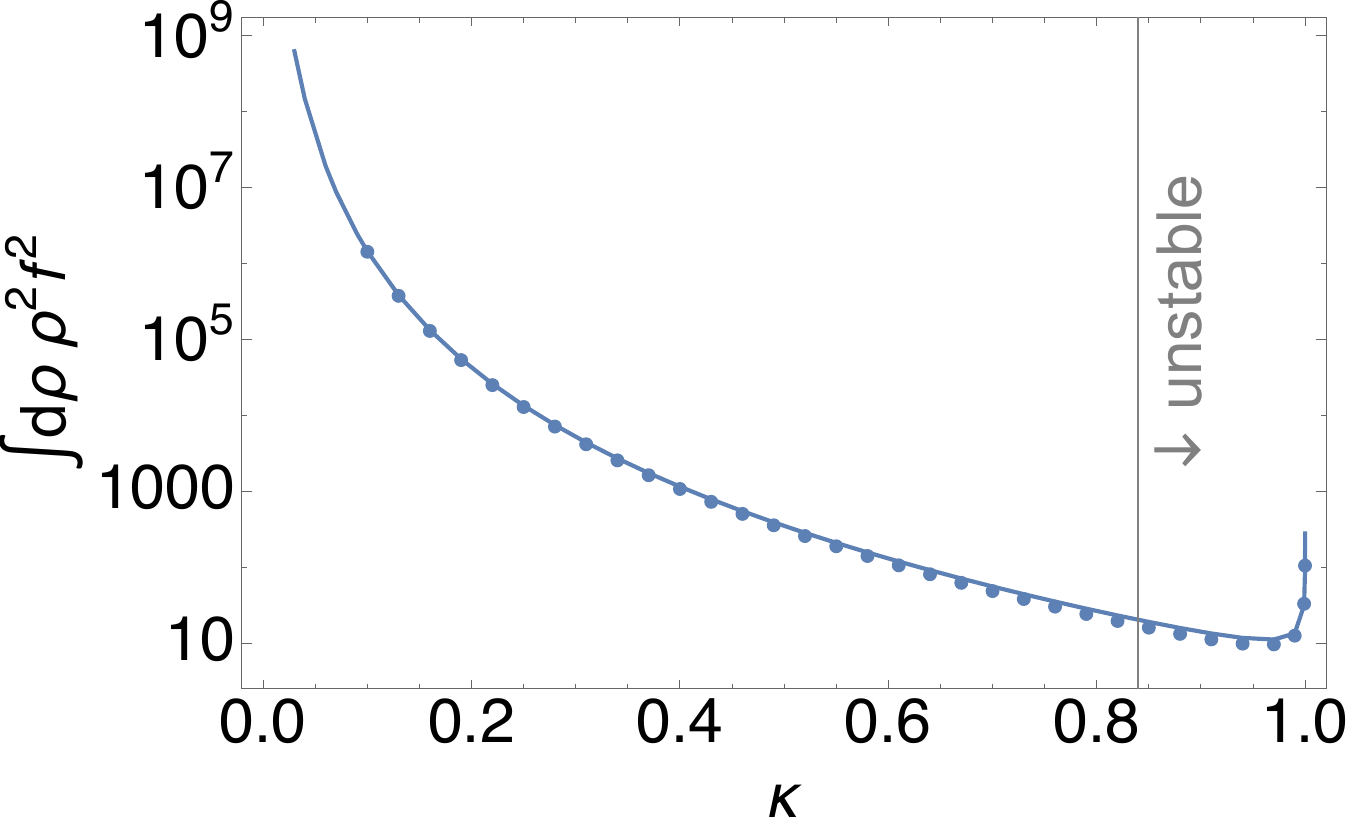}
  \end{subfigure}
  \hfill
  \begin{subfigure}{0.48\textwidth}
    \includegraphics[width=\linewidth]{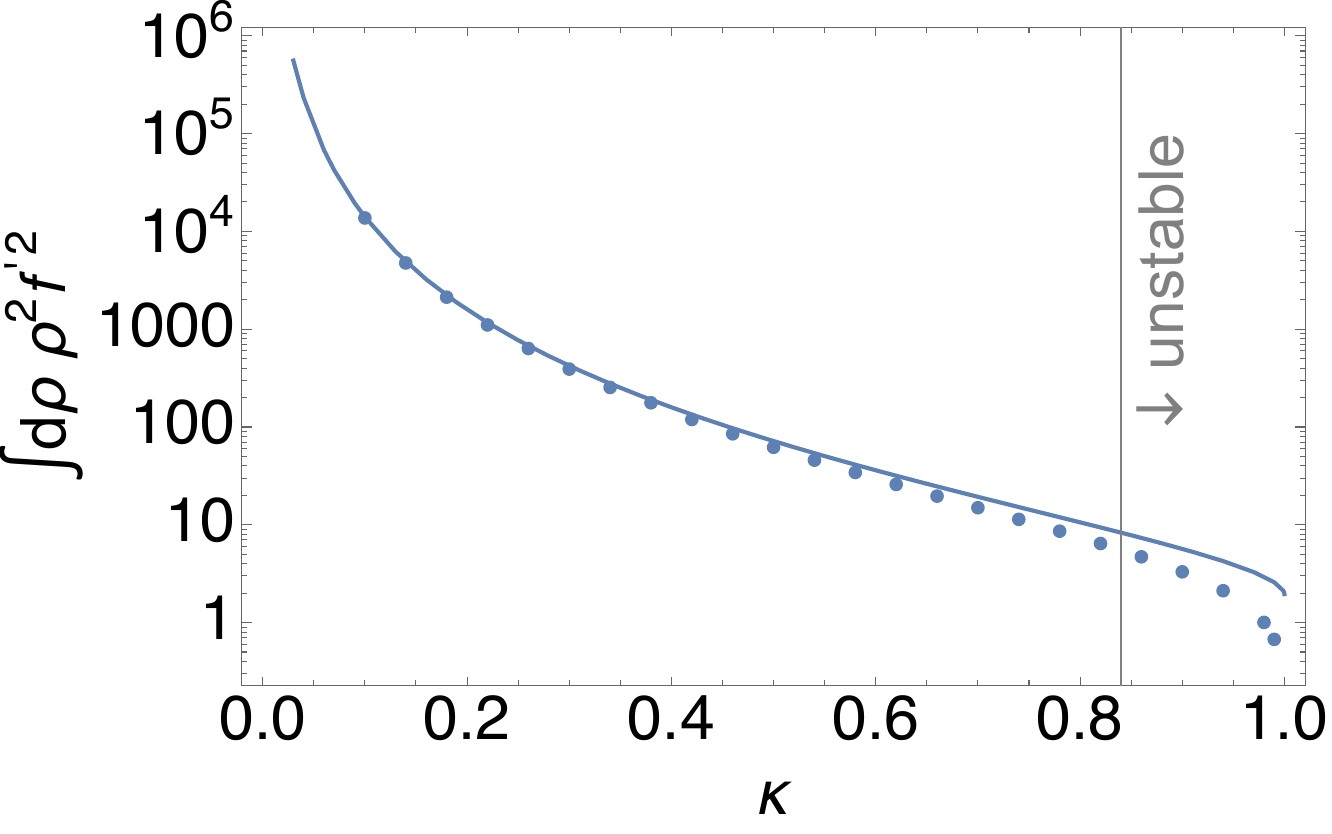}
  \end{subfigure}
  \caption{The integrals relevant for Q-ball charge and energy, $\int \text{d}\rho \rho^2f^2$ (left) and $\int \text{d}\rho \rho^2f^{\prime 2}$ (right), as functions of $\kappa$. The solid lines are the analytical predictions from Eq. (\ref{QintG}) and Eq. (\ref{EintG}), points are numerical results. The region $\kappa > 0.84$ leads to unstable Q-balls.
  }
  \label{fig:QEintGlobal}
\end{figure}

We also show the two integrals that determine $Q$ and $E$ in Fig.~\ref{fig:QEintGlobal}, together with the approximation from Eqs.~\eqref{QintG} and~\eqref{EintG}, which are significantly \emph{better} than the radius prediction. For large $Q$, this leads to the analytic approximations
\begin{align}
    &Q(R) = \frac{\Lambda^2}{m_\phi^2} \left( \frac{81}{4}R^4-27R^3+9R^2+\mathcal{O}(R) \right),\\
    &E(R) = \frac{\Lambda^2}{m_\phi} \left( 18\pi R^3 - 27\pi R^2+\mathcal{O}(R)\right).
\end{align}
By eliminating $R$ or $\kappa$, we can also find the direct relationship between Q-ball energy and charge:
\begin{align}
    E(Q) &= \frac{4\pi}{3}\sqrt{2m_\phi \Lambda}Q^{3/4}-2\pi \Lambda Q^{1/2} +\frac{\sqrt{2} \pi  \Lambda ^{3/2}}{\sqrt{ m_\phi}}Q^{1/4} + \mathcal{O}(Q^{0})\\
    &\simeq m_\phi Q \left[\frac{4\pi\sqrt2}{3} \left(\frac{\Lambda^2/m_\phi^2}{Q}\right)^{1/4} -2\pi \left(\frac{\Lambda^2/m_\phi^2}{Q}\right)^{1/2} +\sqrt2 \pi  \left(\frac{\Lambda^2/m_\phi^2}{Q}\right)^{3/4} \right] .
    \end{align}
We find that these solitons are stable (i.e.~satisfy $E < m_{\phi}Q$) for $\kappa < 0.84$~\cite{Espinosa:2023osv}. We confirm the characteristic flat-potential large-Q-ball behavior $Q\propto \omega^{-4}$, $E\propto Q^{3/4}$~\cite{Dvali:1997qv}, which is markedly different from Coleman's Q-balls, where $E\propto Q$ for large $Q$. Flat-potential solitons hence have a larger binding energy than Coleman's, i.e.~a more attractive force that binds them together. Once the charge grows beyond some critical value, $Q_\text{crit} \simeq 500\, \Lambda^2/m_\phi^2$~\cite{Espinosa:2023osv}, these solitons inevitably become stable and only become more tightly bound as they grow.

\section{Gauged Q-Balls}
\label{sec:gauged}

Promoting the global $U(1)$ symmetry of $\L$ to a \textit{local} symmetry leads to gauged Q-balls~\cite{Rosen:1968zwl,Lee:1988ag}. The additional repulsive long-range interactions mediated by the gauge boson $A_\mu$ make the bound-state construction more difficult. For small Q-balls, this effect is weak, but large Q-balls have a large gauge potential and the resulting Q-balls are qualitatively different from the global case. This has been discussed at length for Coleman's Q-balls, where it has been established that gauged Q-balls cannot grow past some maximal size~\cite{Lee:1988ag,Heeck:2021zvk}. For \textit{flat} potentials, this conclusion might not hold, seeing as the global Q-balls have parametrically larger binding energies than in non-flat potentials. The discussion below confirms the results of Ref.~\cite{Brihaye:2014gua}, however, that gauged Q-balls have a maximal size even in flat potentials.

The Lagrangian now contains the gauge-boson kinetic term and interaction with $\phi$ through the covariant derivative $D_\mu = \partial_\mu - \ii e A_\mu$:
\begin{align}
    \L = |D_\mu \phi|^2 - U(|\phi|) -\tfrac14 F_{\mu\nu}F^{\mu\nu} \,.
\end{align}
We make the usual static-charge ansatz~\cite{Lee:1988ag} for the gauge potential,
\begin{align}
    A_{0} (t,\vec{x}) = \sqrt2\Lambda A(r) \,, \quad A_{1,2,3}(t,\vec{x})=0\,,
\end{align}
with dimensionless gauge profile $A$, and also define the following non-negative parameter that determines the strength of the gauge interactions:
\begin{align*}
       \alpha \equiv e  \frac{\sqrt{2}\Lambda}{m_\phi} \,.
\end{align*}
Everything else, including the scalar potential $U$, is defined as in the global case in Sec.~\ref{sec:global}.

\begin{figure}[t]
  \centering

  \begin{subfigure}{0.48\textwidth}
    \includegraphics[width=\linewidth]{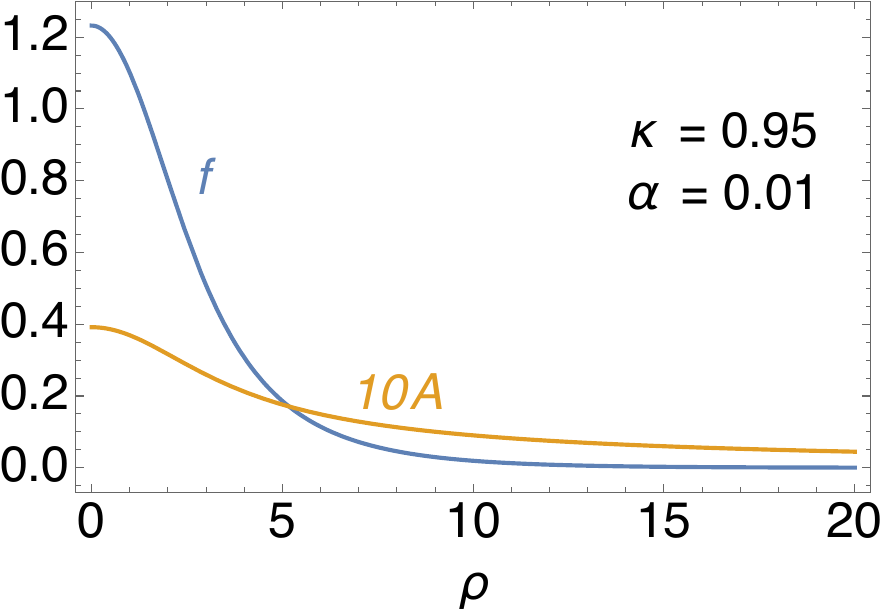}
  \end{subfigure}
  \hfill
  \begin{subfigure}{0.48\textwidth}
    \includegraphics[width=\linewidth]{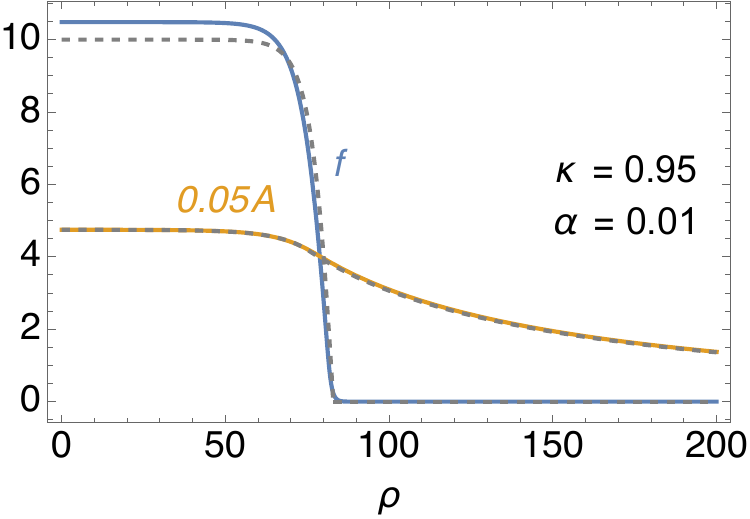}
  \end{subfigure}

  \caption{Profiles for the scalar field $f$ and gauge field $A$ for a small Q-ball (left) and a large Q-ball (right). Solid lines are numerical solutions. Dashed lines are analytical predictions from Eq.~\eqref{Apred} and Eq.~\eqref{fpred} for large Q-balls.}
  \label{fig:fGauged}
\end{figure}

The potential in which the particle $f(\rho)$ rolls now becomes $\rho$ (or ``time'') dependent through $A(\rho)$, leading to qualitatively different solutions:
\begin{align}
    V (f,A) 
    & = \frac{1}{2}f^2\left(\kappa - \alpha A\right)^2 - \frac{1}{2}\left(1 - e^{-f^2}\right).
\end{align}
The differential equations for $f$ and $A$ are
\begin{align}
  &f^{\prime\prime} + \frac{2}{\rho}f^\prime = fe^{-f^2} - \left(\kappa - \alpha A\right)^2 f, \label{eq:gauged_f} \\
    &A^{\prime\prime} + \frac{2}{\rho}A^\prime = \alpha f^2 \left(A \alpha - \kappa\right) ,
\end{align}
with the same boundary conditions for $f$ and $A$; $A$ is monotonically decreasing, too. 
Numerical solutions are again obtained by switching from $\rho$ to $y$ and using appropriate test functions for our finite-elements method. Two examples are shown in Fig.~\ref{fig:fGauged}. Just like gauged Q-balls in non-flat potentials~\cite{Heeck:2021zvk}, more than one solution can exist for the same parameter point, typically split into a branch that resembles small global Q-balls [Fig.~\ref{fig:fGauged} (left)] and a branch of large Q-balls with significantly larger gauge field $A$ [Fig.~\ref{fig:fGauged} (right)].

\begin{figure}[tb]
    \centering
    \includegraphics[width=0.7\textwidth]{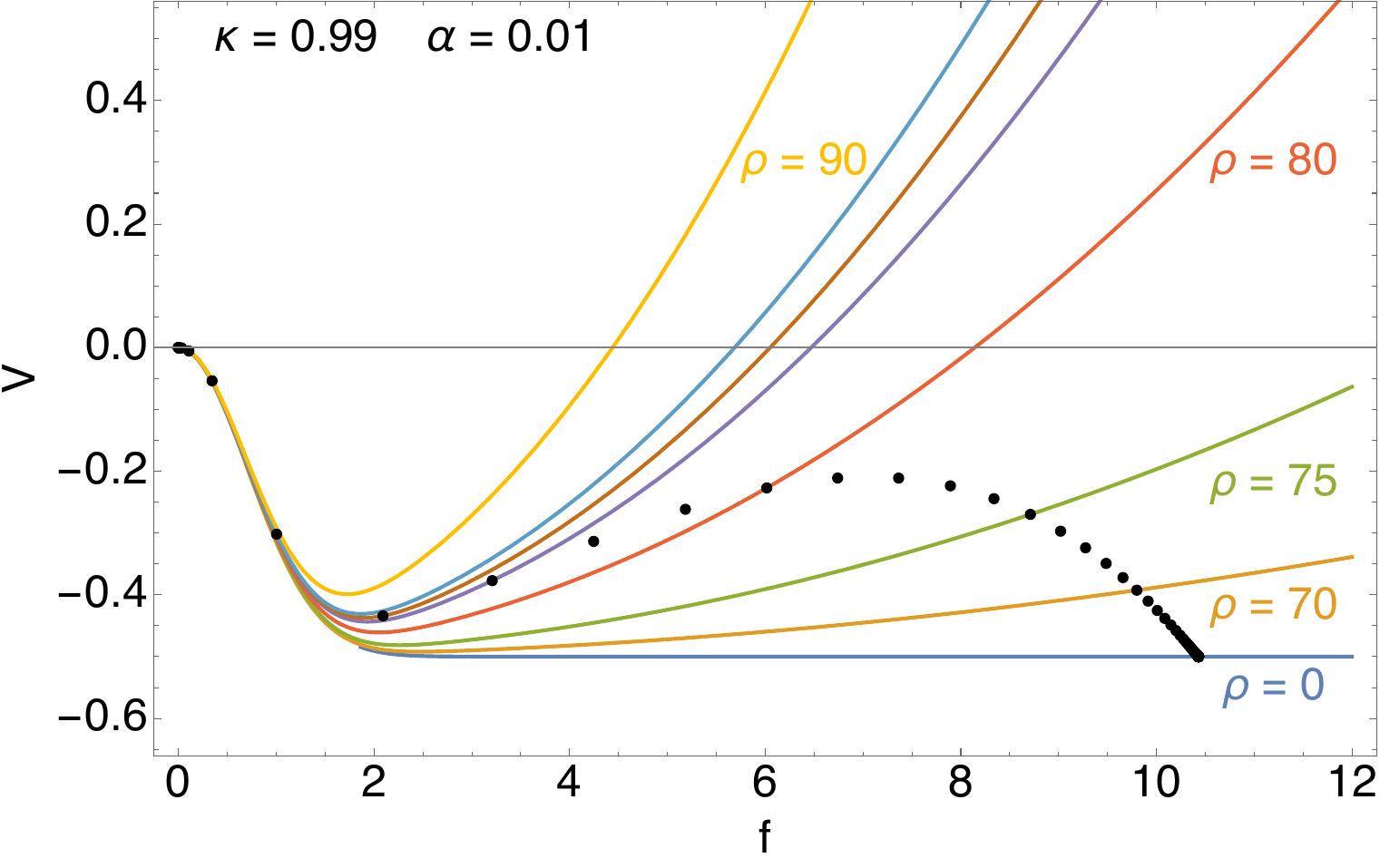}
    \caption{Effective potential of $f$ given a specific $A(\rho)$. Black points indicate the values of the gauged Q-ball profile within a range $\rho \in [0,100]$.}
     \label{fig:VfGauged}
\end{figure}

For $\alpha A\ll \kappa$, the small Q-ball will resemble the global one, but interesting effects arise for $\alpha A\sim \kappa$. In the extreme case of $\alpha A(0)\simeq \kappa$ and $f\gg 1$, the right-hand side of Eq.~\eqref{eq:gauged_f} is approximately zero, allowing for a constant $f$ solution. Eventually, $f$ and $A$ start dropping to zero, but the core of the Q-ball has a nearly constant $f$ and $A$ density, just like in non-flat potentials!
The appropriate motion in the effective potential is shown in Fig.~\ref{fig:VfGauged}: particle $f$ starts off on an almost horizontal potential, i.e.~force free, and then starts rolling when the potential starts tilting, which supplies the energy needed to make it to the maximum at $f=0$.

The scalar field solution for these large gauged Q-balls can then be approximated as a Heaviside step function with some initial value $f(0)$, $f = f(0)(1 - \Theta(\rho - R))$, where $R$ is the radius of the Q-ball. This simple ansatz allows us to solve the $A$ equation as
\begin{equation}
    A\left(\rho\right) = \frac{\kappa}{\alpha} 
\begin{cases}
1 - \frac{\sinh(\alpha f(0) \rho)}{\alpha f(0) \rho \cosh(\alpha f(0) R)}\,, & \rho < R \,,\\[6pt]
\frac{\alpha R - \tanh\left(\alpha f(0) R\right)/f(0)}{\alpha \rho}\,, & \rho \geq R\,,
\end{cases}
\label{Apred}
\end{equation}
which is a fantastic approximation for large Q-balls, just like in non-flat potentials~\cite{Lee:1988ag,Heeck:2021zvk}. 
Throwing this expression for $A$ back into the $f$ equation and approximating $f\gg 1$ inside the Q-ball, we find,
\begin{align}
    f^{\prime\prime} + \frac{2}{\rho}f^\prime + \left(\frac{\kappa\sinh(f(0) \alpha \rho) }{f(0) R \alpha \cosh(f(0) \alpha R)}\right)^2 f \simeq 0\,,
\end{align}
which can be solved by a modified odd Mathieu function. Further approximating $\sinh(x)\simeq \exp(x)/2$ yields a more familiar Bessel-type differential equation, whose solution can be approximated by
\begin{equation}
f(\rho)\simeq
\begin{cases}
f(0)\, J_0\!\left(e^{\,f(0) \alpha (\rho - R)}\right),
  & 0 \le \rho \le R + \dfrac{\ln j_{0,1}}{f(0) \alpha}\,,\\[6pt]
0\,,
  & \rho \ge R + \dfrac{\ln j_{0,1}}{f(0) \alpha}\,,
\end{cases}
\label{fpred}
\end{equation}
where $j_{0,1}\simeq 2.405$ is the first positive zero of $J_0$. Strictly speaking, this approximation is only valid inside the Q-ball, i.e.~for $\rho < R$, but extending it a bit further to $R + \dfrac{\ln j_{0,1}}{f(0) \alpha}$ gives a decent and simple approximation for the entire profile that improves upon the initial step function, see Fig.~\ref{fig:fGauged}.

Further study of the differential equations as well as work--friction relationships~\cite{Heeck:2021zvk} in the large Q-ball limit lead to the following approximations for $f(0)$ and $R$:
\begin{align}
    f(0) \simeq \frac{1}{\sqrt{\alpha}}\,, && 
    R \simeq \left(\frac{\kappa}{\alpha} - \frac{4f(0) \alpha}{3}\right).
    \label{LQR}
\end{align}
The analytic expressions for $f$ and $A$ are surprisingly good on the large-Q-ball branch, as can be seen in Fig.~\ref{fig:fGauged} (right).
Since $\kappa \leq 1$ even for gauged Q-balls, the radius approximation implies a maximal radius of $1/\alpha$. Despite the large qualitative difference between flat and non-flat \textit{global} Q-balls, the \textit{gauged} versions are remarkably similar at the macroscopic level.

\begin{figure}[tbp]
  \centering

  \begin{subfigure}{0.48\textwidth}
    \includegraphics[width=\linewidth]{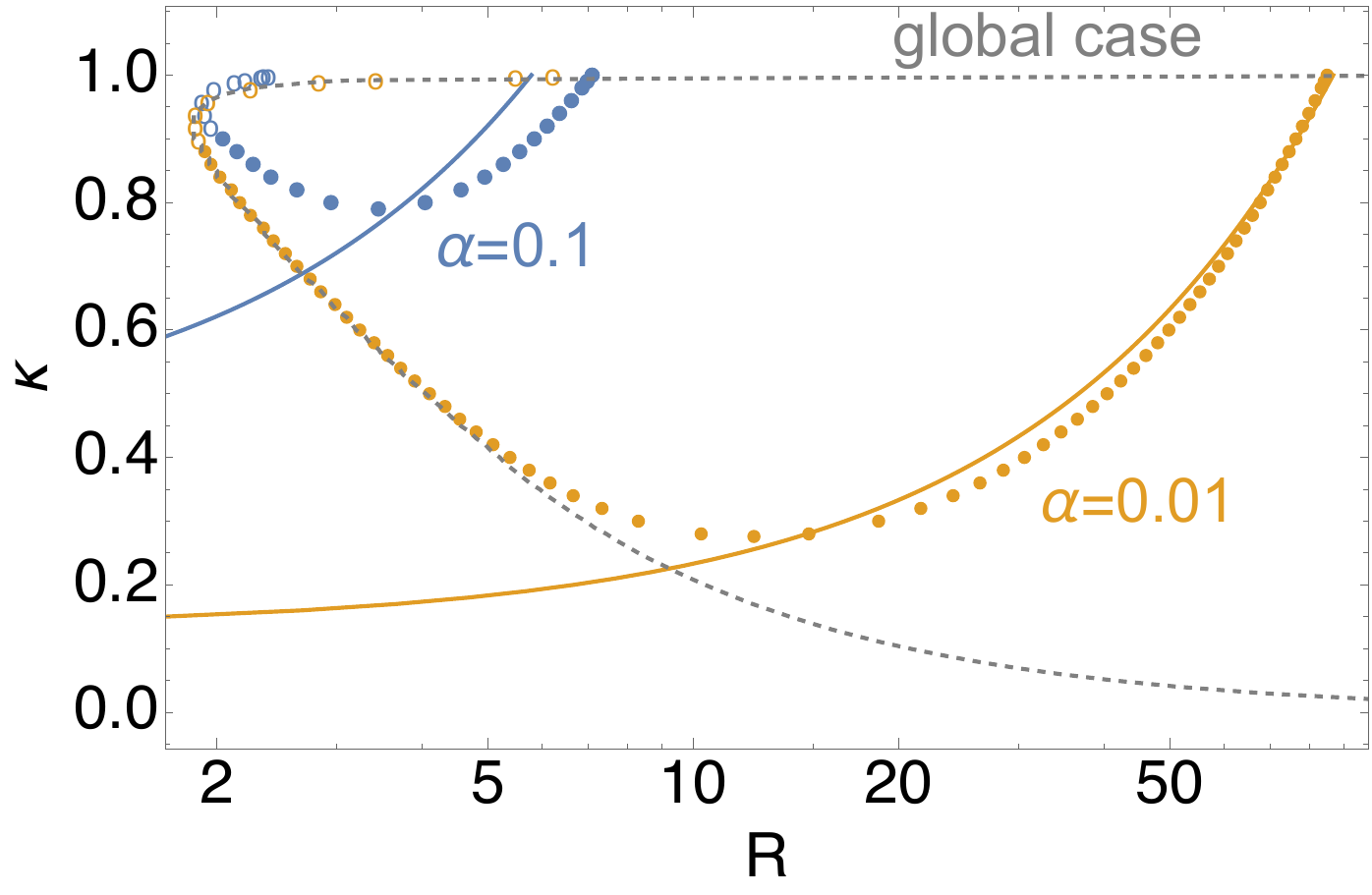}
    \label{fig:subfig1}
  \end{subfigure}
  \hfill
  \begin{subfigure}{0.48\textwidth}
    \includegraphics[width=\linewidth]{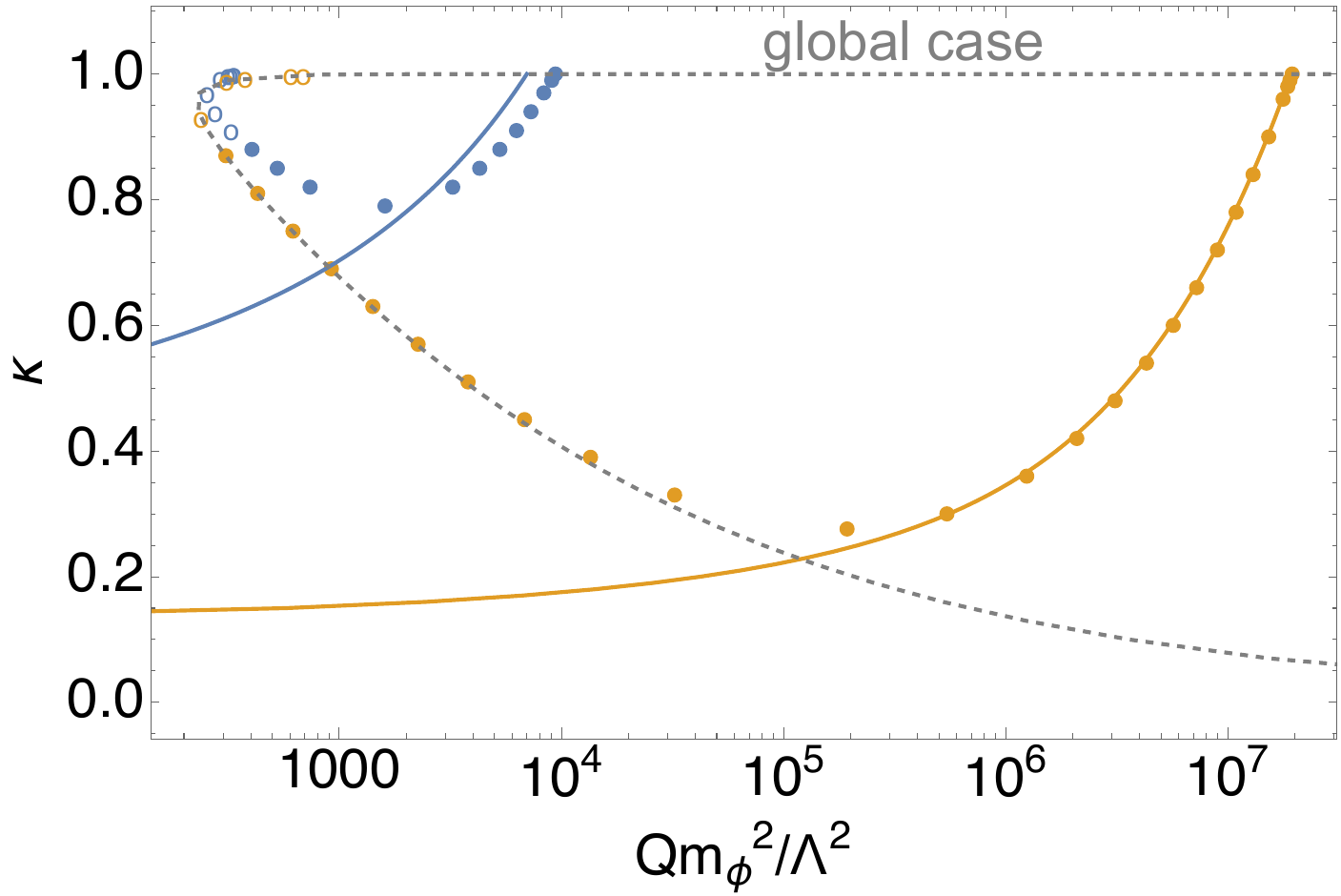}
    \label{fig:subfig2}
  \end{subfigure}

  \begin{subfigure}{0.48\textwidth}
    \includegraphics[width=\linewidth]{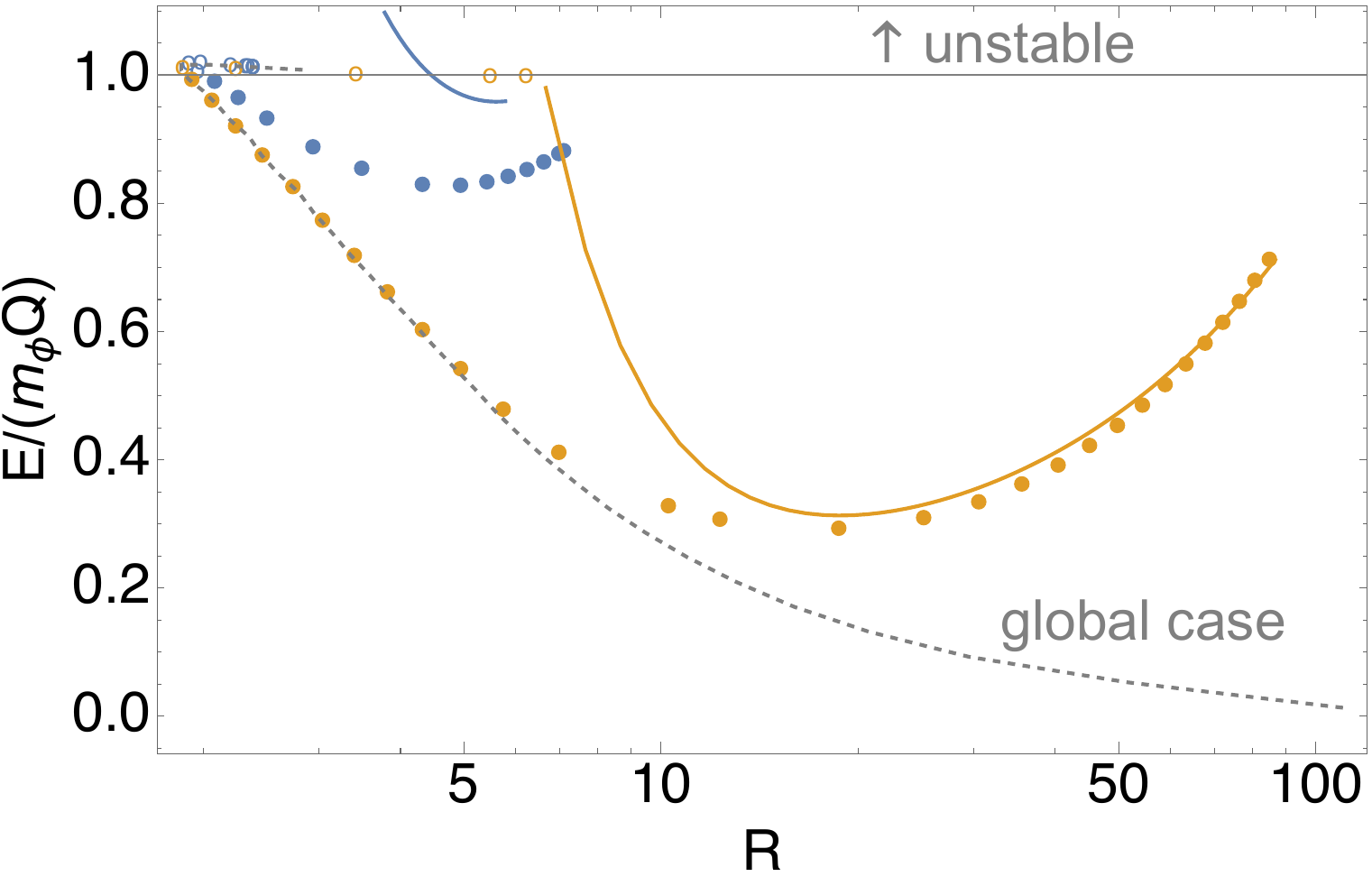}
    \label{fig:subfig3}
  \end{subfigure}
  \hfill
  \begin{subfigure}{0.48\textwidth}
    \includegraphics[width=\linewidth]{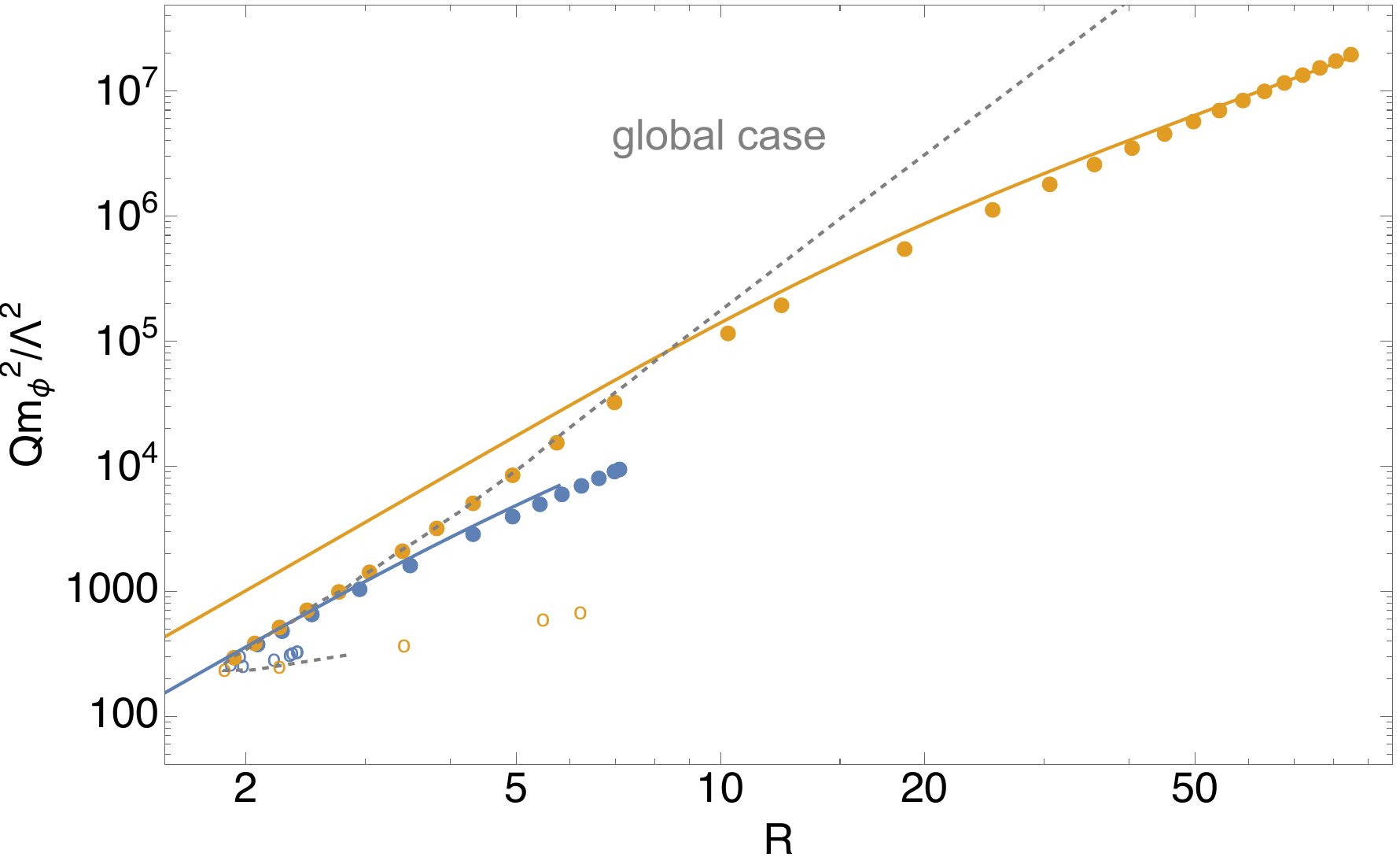}
    \label{fig:subfig4}
  \end{subfigure}

  \caption{Points are numerical solutions for benchmarks $\alpha = 0.01$ (orange) and $\alpha = 0.1$ (blue); open points indicate unstable cases ($E>m_\phi Q$). The global Q-ball case is shown in dotted gray for comparison. Our analytic large Q-ball predictions are shown as the solid lines.}
  \label{fig:4graphGauged}

\end{figure}

We can use these approximations to estimate the charge
\begin{align}
    Q & = 8 \pi \frac{\Lambda^2}{m_\phi^2} \int \text{d}\rho \rho^2 f^2 \left(\kappa - \alpha A\right) 
\end{align}
and energy
\begin{align}
    E/m_\phi & = \kappa Q + \frac{8  \pi \Lambda^2}{3m_\phi^2}\int \text{d}\rho \rho^2 \left(f^{'2} - A^{'2}\right) 
\end{align}
 of the large gauged Q-balls in our flat potential. The relevant integrals take the form
 \begin{align}
    &\int \text{d}\rho \rho^2 f^2\left(\kappa - \alpha A\right) = \frac{\kappa \left(f(0) R \alpha - \tanh (f(0) R \alpha) \right)}{f(0) \alpha^3} \,,
    \label{Qint}\\
    &\int \text{d}\rho \rho^2 \left(f^{'2} - A^{'2}\right) = -\frac{f(0)R^3\alpha^3 - 6\kappa^2 + 3f(0)R\alpha\kappa^2}{4f(0) 
    \alpha^3} \,.
    \label{Eint}
\end{align}
We show the macroscopic Q-ball quantities -- radius, charge, and energy -- in Fig.~\ref{fig:4graphGauged} for two benchmark values of $\alpha$,\footnote{The numerical solutions are difficult to obtain even using our analytical expressions as test functions. Instead, we start with the solutions of $\kappa = 0.8$, $\alpha = 0.1$ and use these functions as the guess functions to obtain the solutions of the next $\kappa = 0.8+\epsilon$ or $\alpha = 0.1+\epsilon$ value.  } together with our large Q-ball approximations derived above.

\begin{figure}[tbh]
    \centering
    \includegraphics[width=0.6\textwidth]{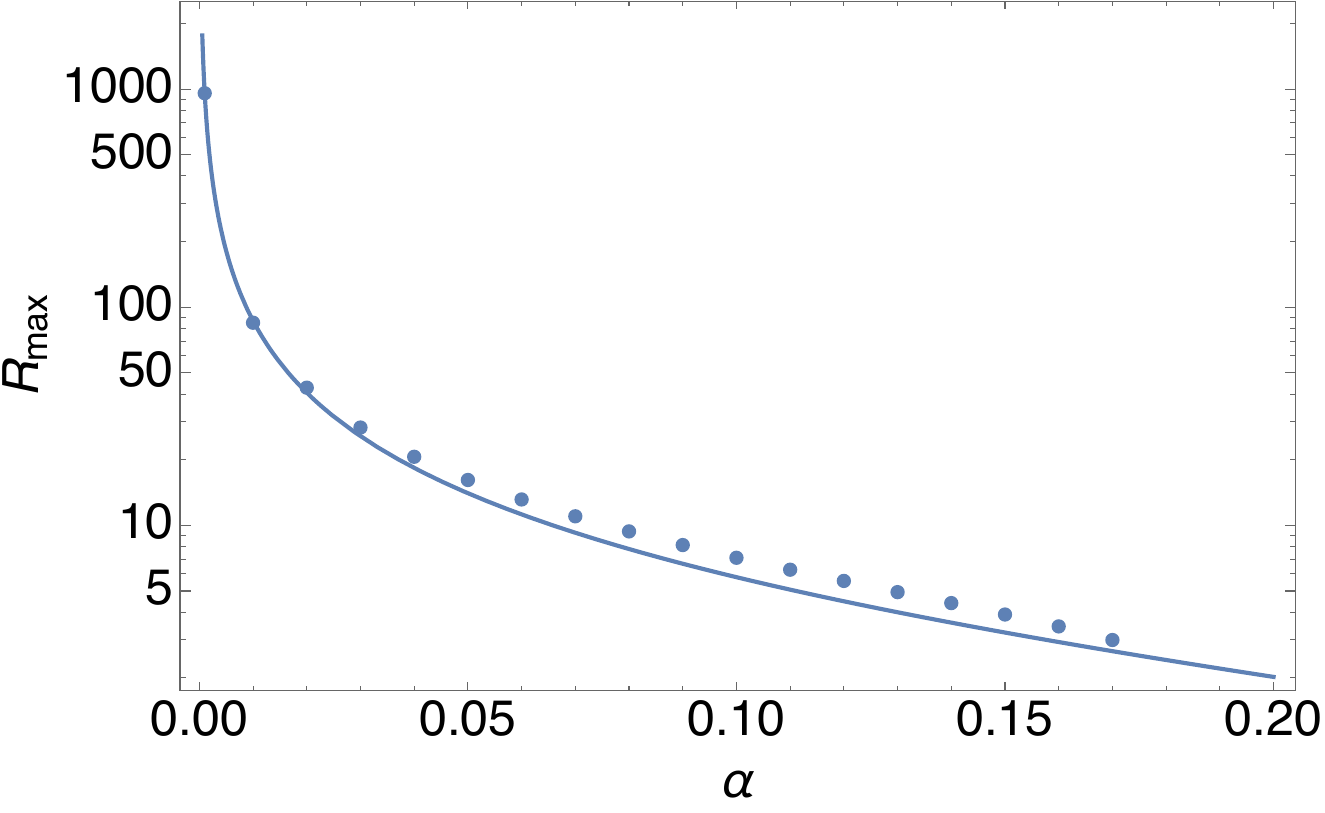}
    \caption{Maximal radius $R$ vs.~$\alpha$ as points; the solid line is the $\kappa = 1$ prediction of Eq.~\eqref{LQR}. There are no Q-ball solutions for $\alpha > 0.17$.}
     \label{fig:RamxGauged}
\end{figure}

Fig.~\ref{fig:4graphGauged} illustrates and confirms some basic qualitative properties of gauged Q-balls in flat potentials: for small $\alpha$, or weak gauge interactions, small Q-balls closely resemble the global case from Sec.~\ref{sec:global}; once the radius grows beyond $1/\sqrt{\alpha}$, the gauge repulsion starts to kick in and leads to the qualitatively different large thin-wall Q-balls, which can grow up to a radius of $1/\alpha$ [see Fig.~\ref{fig:RamxGauged}], or a charge of $Q_\text{max} \simeq 8\pi \Lambda^2/(m_\phi^2 \alpha^3) = \sqrt{8} \pi m_\phi/(\Lambda e^3)$. For small $\alpha$, that branch is well approximated by our analytical expressions, which however fail for smaller Q-balls. 
For $\alpha > 0.17$, no Q-ball solutions exist, in agreement with Ref.~\cite{Brihaye:2014gua}. Macroscopically, gauged Q-balls in flat potentials look remarkably similar to gauged Q-balls in non-flat potentials, albeit more difficult to approximate since there is no mapping relation~\cite{Heeck:2021zvk}.

For large global Q-balls, our potential~\eqref{eq:Uexact} furnishes an excellent approximation for other flat potentials, such as the supersymmetry-inspired logarithmic potential discussed in Refs.~\cite{Hong:2015wga,Campanelli:2007um}. In the gauged case, the degree of flatness plays a role though: the step-function approximation for $f(\rho)$ from Eq.~\eqref{fpred} hinges on the flatness of $V$ near $\rho = 0$, see Fig.~\ref{fig:VfGauged}. For potentials with a logarithmic flatness, $f$ will start rolling earlier and not be constant inside the Q-ball, which leads to a different $Q_\text{max} \propto 1/e^3$ scaling for the maximal charge~\cite{Hong:2015wga}. The qualitative result of a maximal charge persists, however.

\clearpage

\section{Proca Q-Balls}
\label{sec:proca}

If the gauge boson $A$ is not massless but rather has a mass $m_A \equiv m_\phi M$, the Q-ball solutions change yet again. For small $M$, the Q-ball will look like a gauged Q-ball, while large $M$ basically decouples the gauge boson and yields an effectively global Q-ball. In the intermediate region, new effects occur~\cite{Heeck:2021bce}. The differential equations now contain an extra term $M^2 A$:
\begin{align}
    &f^{\prime\prime} + \frac{2}{\rho}f^\prime = fe^{-f^2} - \left(\kappa - \alpha A\right)^2 f \,,\label{eq:ProcafEq} \\
    &A^{\prime\prime} + \frac{2}{\rho}A^\prime = \alpha f^2 \left(A \alpha - \kappa\right) + M^2 A\,.
    \label{eq:ProcaAEq}
\end{align}
Just like in the gauged case, there are between zero and two solutions for a given $\kappa$, unlike in non-flat potentials which could have even more solutions depending on $M$~\cite{Heeck:2021bce}. 

\begin{figure}[tbh]
    \centering
    \includegraphics[width=0.65\textwidth]{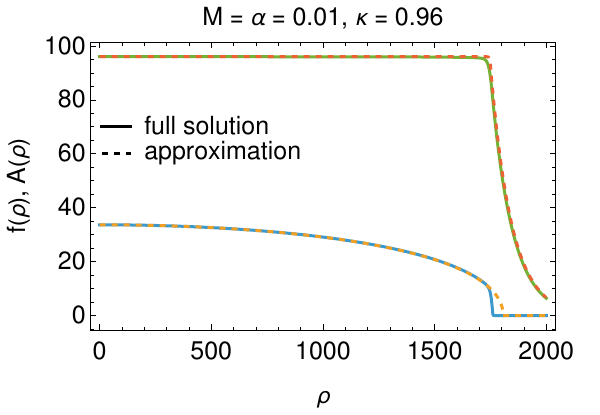}
    \caption{Solutions to the Q-ball differential equations~\eqref{eq:ProcafEq}--\eqref{eq:ProcaAEq} (solid) as well as the large-Q-ball approximations from Eq.~\eqref{eq:AstepProca} and~\eqref{eq:ProcafEq2}, using Eq.~\eqref{eq:radius_approx} for the radius together with the numerical $f(0)$. }
     \label{fig:proca_profile}
\end{figure}

For radii $R\ll 1/\sqrt{\alpha}$, the Q-ball charge is small enough to render these objects indistinguishable from the global case, so only $R>1/\sqrt{\alpha}$ is sensitive to $M$.
The solutions for $f(\rho)$ in the Proca case are different from both the global and gauged case, again unlike the situation in non-flat potentials: $f(\rho)$ is no longer approximately constant inside the Q-ball. However, $A(\rho)$ still is almost constant inside, and is surprisingly well described by the solution one would obtain for a step-function source $f(\rho)$:
\begin{equation}
    A\left(\rho\right) = \frac{\alpha\kappa f(0)^2 }{\mu^2} 
\begin{cases}
1 - \frac{(1+M R) }{\left(M \sinh \left(R \mu\right)+\mu \cosh \left(R \mu\right)\right)  }\, \frac{\sinh \left(\rho \mu\right)}{ \rho}\,, & \rho < R \,,\\[6pt]
e^{M R}\frac{R\mu - \tanh\left(R\mu \right)}{\mu + M\tanh\left(R\mu \right)}\, \frac{e^{-M\rho}}{\rho}\,, & \rho \geq R\,,
\end{cases}
\label{eq:AstepProca}
\end{equation}
with gauge boson mass in the Q-ball center $\mu \equiv \sqrt{\alpha ^2 f(0)^2+M^2}$~\cite{Heeck:2021bce}. More accurately, a constant $A$ inside the Q-ball requires the right-hand side of Eq.~\eqref{eq:ProcaAEq} to vanish, which gives $A(\rho) = f(\rho)^2 \alpha\kappa/(M^2 + f(\rho)^2\alpha^2)$. Throwing this back into Eq.~\eqref{eq:ProcafEq} and assuming $f(\rho) \gg \max(1,M/\alpha)$ inside gives an approximate differential equation for $f$:
\begin{align}
    &f^{\prime\prime} + \frac{2}{\rho}f^\prime +\frac{ \kappa^2 M^4/\alpha^4}{f^3}  =0\,.\label{eq:ProcafEq2}
\end{align}
The solution can be Taylor-expanded as
\begin{align}
    f(\rho) = f(0) \left(1- \frac{1}{6}\left(\frac{\kappa M^2}{\alpha^2 f(0)^2}\right)^2\rho^2 - \frac{1}{40}\left(\frac{\kappa M^2}{\alpha^2 f(0)^2}\right)^4\rho^4 + \mathcal{O}(\rho^6)\right), 
\end{align}
but a more useful observation is that the ansatz $f(\rho ) = f(0) \, u[\rho\,\kappa M^2/(\alpha^2 f(0)^2)]$ solves Eq.~\eqref{eq:ProcafEq2} if $u(x)$ satisfies the parameterless differential equation
\begin{align}
    u''(x) + \frac{2}{x}u'(x) + \frac{1}{u(x)^3} = 0\,,
\end{align}
with $u(0)=1$ and $u'(0)=0$. This differential equation can easily be solved numerically, with $u(x)$ decreasing monotonically until hitting zero at $x= x^* \simeq 1.53407$. The underlying approximation stops being valid for small $f$, say when $f \sim 10$, so the  actual Q-ball radius is smaller, approximately around
\begin{align}
    R \simeq x^*\frac{\alpha^2 f(0)^2}{\kappa M^2} - \frac{55 \alpha^2 }{\kappa M^2} \,.
    \label{eq:radius_approx}
\end{align}
Despite the crudeness of this approximation, it is in excellent agreement with the full solution of the coupled differential equation, as long as one chooses the correct $f(0)$, see Fig.~\ref{fig:proca_profile}. Notice that the true solution for $f(\rho)$ drops to zero even faster than our approximation, which follows $u(x)\simeq\sqrt2 \sqrt{x^*-x}$ near zero, providing a sharp edge for the Q-ball. This drop is not caused by the exponential term in the potential, which is still heavily suppressed around $f\sim 10$, but rather by $A$ no longer being constant.

\begin{figure}[tbh]
    \centering
    \includegraphics[width=0.9\textwidth]{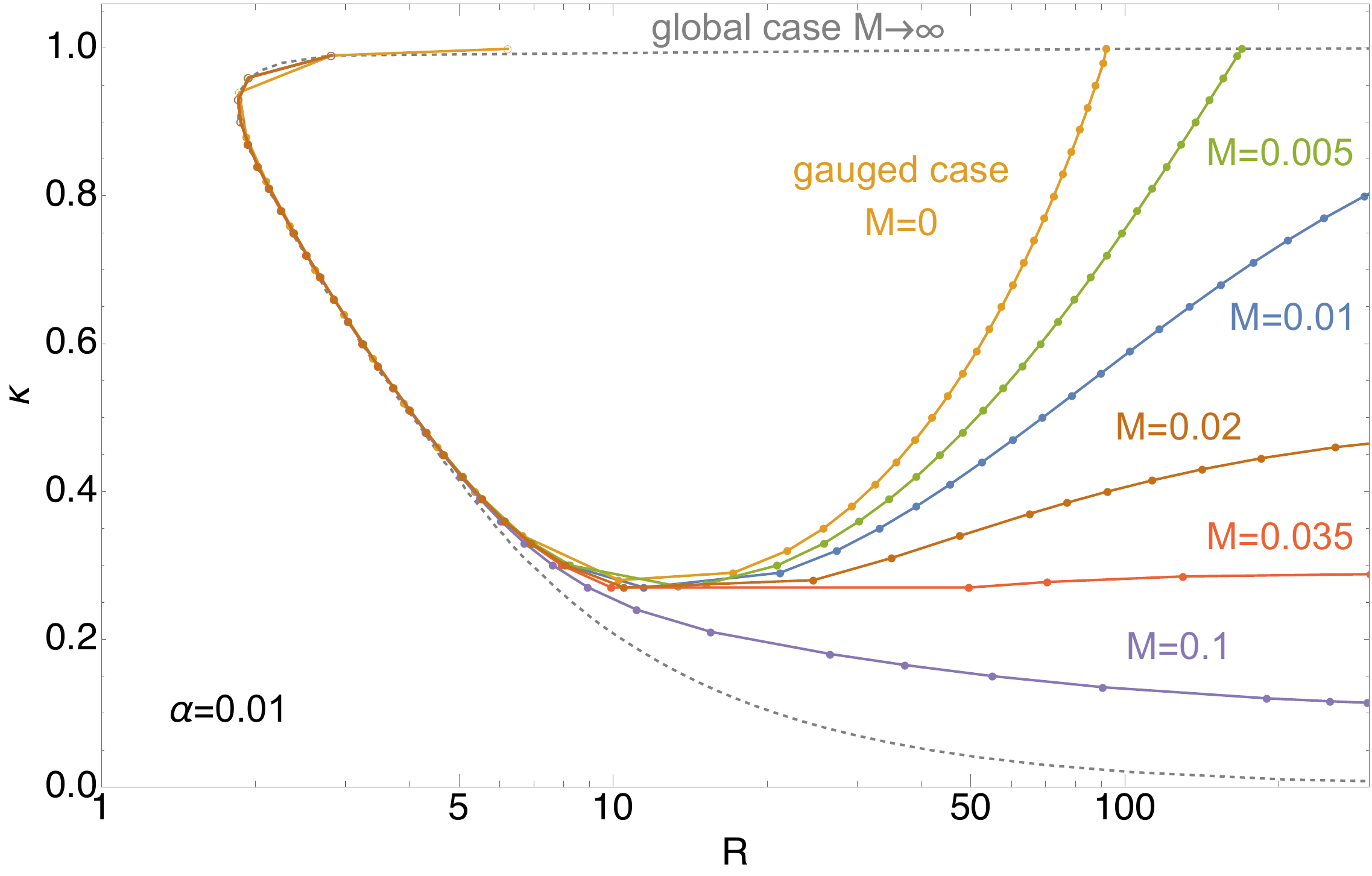}
    \caption{Q-ball radius $R$ vs.~$\kappa$ for $\alpha = 0.01$ and various $M$ values, including the extreme cases $M=0$ (the gauged case, orange) and $M\to \infty$ (the global case, dotted gray). Points are numerical solutions, the lines are \emph{not} analytical predictions but just interpolation to guide the eye. Open points indicate unstable cases ($E>m_\phi Q$) for small Q-balls near $\kappa\sim 1$.}
     \label{fig:ProcaRvsKJoined}
\end{figure}

\begin{figure}[tbp]
  \centering
  \begin{subfigure}{0.48\textwidth}
       \includegraphics[width=\linewidth]{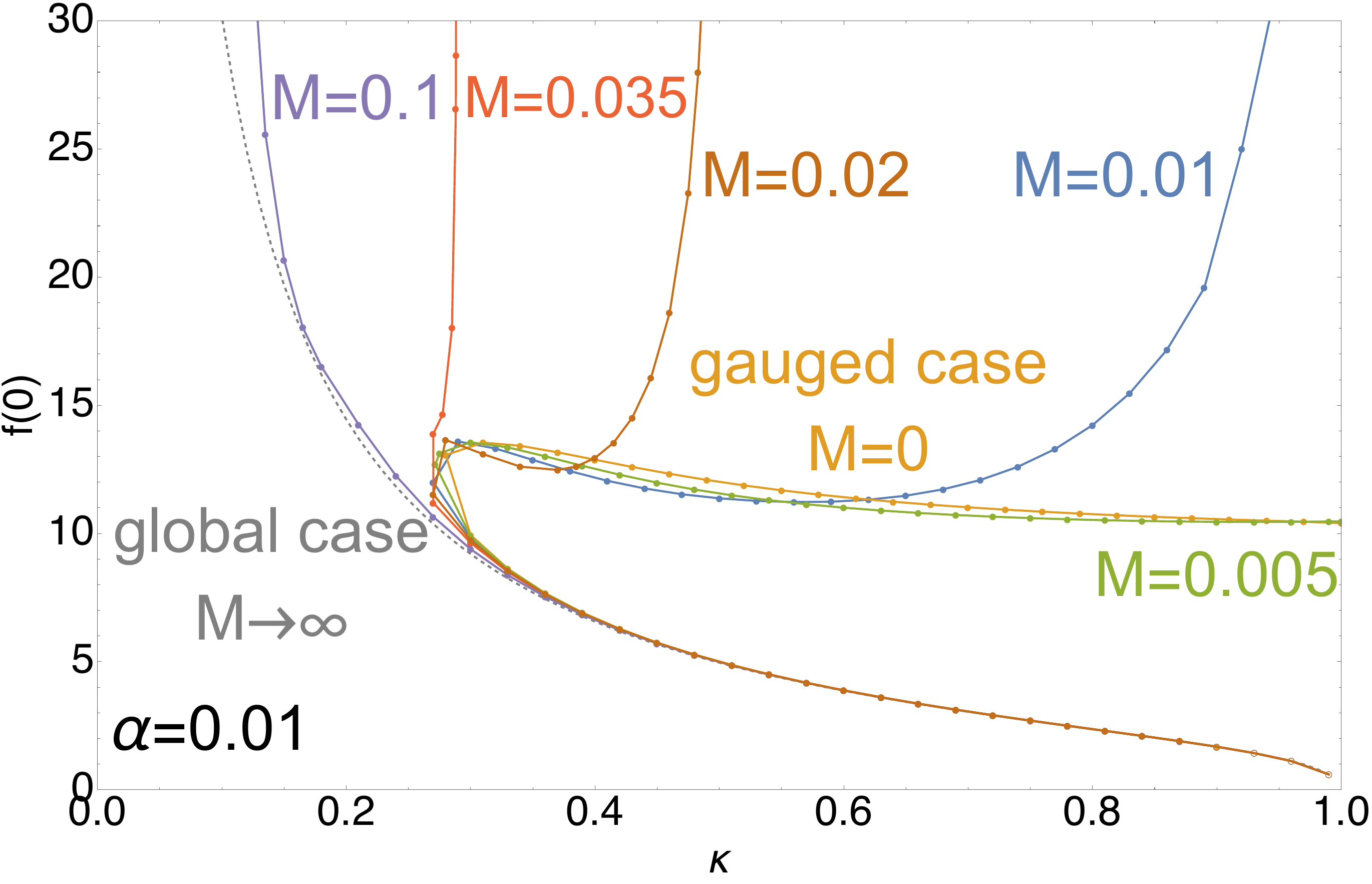}
  \end{subfigure}
\hfill
  \begin{subfigure}{0.48\textwidth}
      \includegraphics[width=\linewidth]{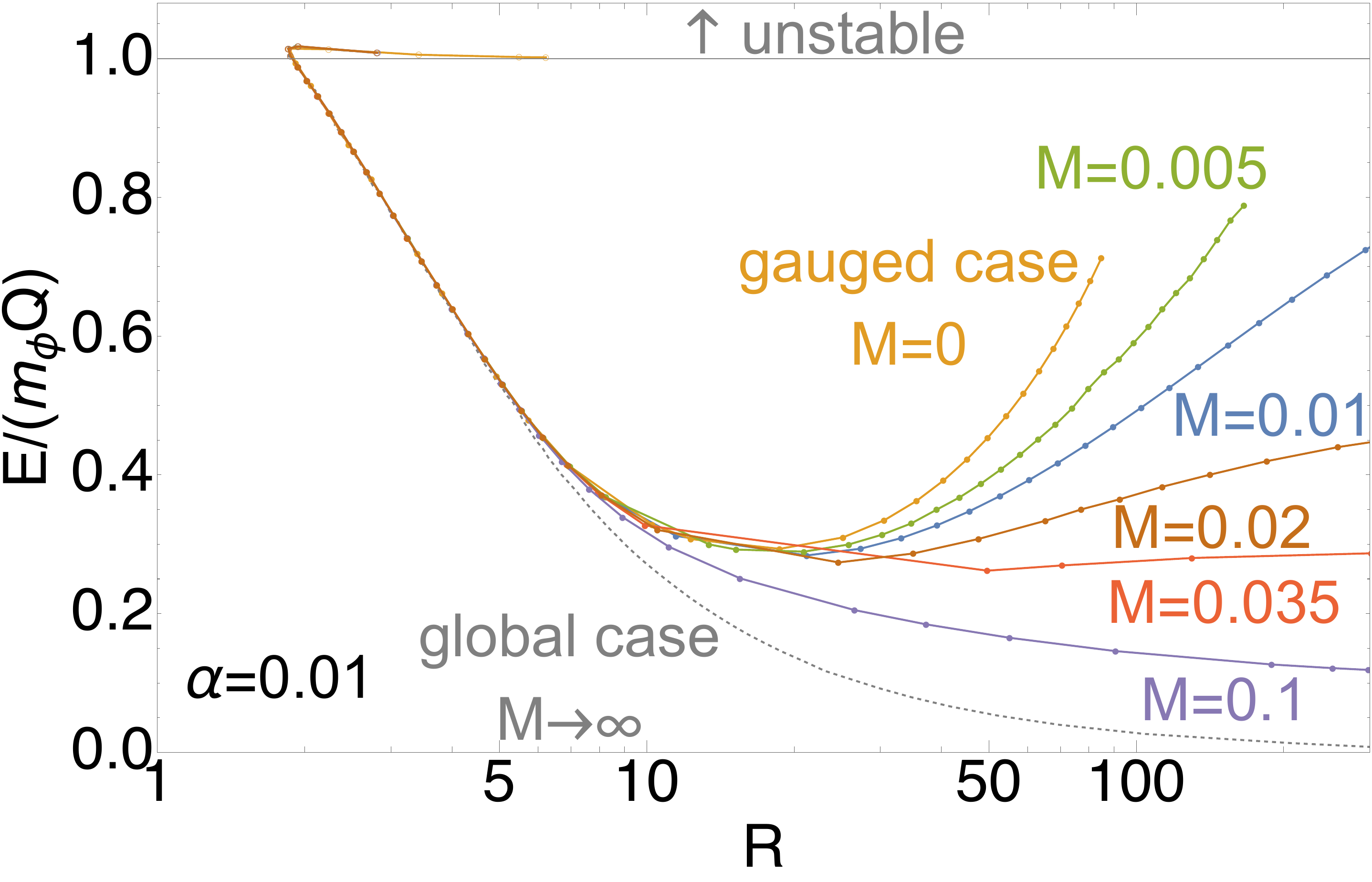}
  \end{subfigure}
  \caption{Similar to Fig.~\ref{fig:ProcaRvsKJoined} but showing $f(0)$ vs.~$\kappa$ (left) and $E/(m_\phi Q)$ vs.~$R$ (right). }
  \label{fig:ProcaEoQandf0}
\end{figure}

The last missing piece is a relationship between $f(0)$ and $\kappa$, which is unfortunately difficult to obtain. By using energy-lost-to-friction arguments~\cite{Heeck:2021bce}, we can at least show that $f(0)$ diverges for $\kappa\to \alpha/M$, and, through Eq.~\eqref{eq:radius_approx}, so does the radius. For $M < \alpha$, this implies a finite maximal Q-ball radius once $\kappa$ hits $1$, whereas Q-balls with $M\geq \alpha$ can grow to arbitrary size/charge, as illustrated in Fig.~\ref{fig:ProcaRvsKJoined}. 
Fig.~\ref{fig:ProcaEoQandf0} (left) shows $f(0)$ vs.~$\kappa$ and how it diverges for $\kappa\to \alpha/M$ if $M\geq\alpha$. $M\simeq 0.035$ marks another interesting boundary: for $M\gtrsim 0.035$, the radius increases for decreasing $\kappa$, just like for global Q-balls, whereas for $M\lesssim 0.035$, $\kappa (R)$ has a minimum around $R\sim 1/\sqrt{\alpha}$.
All of these large Q-balls are stable in the sense that $E < m_\phi Q$ (see Fig.~\ref{fig:ProcaEoQandf0}), but since stability has not even been proven for the gauged case in non-flat potentials, we will refrain from claiming absolute stability for these large Proca Q-balls, to be revisited in future work.

\clearpage

\section{Conclusions}
\label{sec:conclusions}

Solitons in flat potentials have attracted enormous interest over the last decades once it was realized that they generically exist in supersymmetric versions of the Standard Model and could form dark matter. These are typically approximated as single-field problems, neglecting any and all gauge interactions. In this article, we have for the first time explored gauged and Proca Q-balls in flat potentials, paving the way to a more realistic description of these objects in the presence of additional forces. Global, gauged, and Proca Q-balls all exhibit different properties, unlike in non-flat potentials where the solutions are related through mapping. Importantly, gauged Q-balls in flat potentials exhibit a maximal radius just like in non-flat potentials, precluding further growth of these solitons. The same occurs for Proca Q-balls if the gauge boson mass is below some critical value, otherwise the solitons resemble the global case. Following this initial exploration, we aim to study more realistic supersymmetric Q-balls in future work.

\section*{Acknowledgements}

JH thanks Christopher B.~Verhaaren and Arvind Rajaraman for discussions during the early stages of this project.
This work was supported by the U.S. Department of Energy under Grant No.~DE-SC0007974.
Data tables of our numerical solutions are available as ancillary files of the arXiv version of this article~\cite{Heeck:2026iqe}.

 \bibliographystyle{JHEP}

\end{document}